\documentstyle[preprint,aps]{revtex}
\tighten \begin{document}
\draft
\title{On the Collective Mode Spectrum in FQHE}
\author{ Alejandro Cabo and Aurora P\'erez Mart\'{\i}nez}
\address{Group of Theoretical Physics, ICIMAF\\
Calle E $\#$ 309, esq. a 15, CP 10400 Ciudad Habana, Cuba}
\maketitle
\begin{abstract}
We consider the Bethe Salpeter Equation (BSE) for a fractionally
filled Landau level. A phenomenological discussion of the $1/3$
Laughlin's state is performed by assuming an ansatz for the
one-particle propagator.
The BSE is solved in this approach and it predicts
an instability under the formation of charge density oscillations
for  a wide range of the one-particle gap parameter values in
contrast with previous single mode approximation results.
However, the conclusion is  compatible with the one obtained within
a composite fermion description done by us before
and with the saturation of the zero momentum oscillator strength
sum rule by the cyclotronic resonance.
Further studies should be done in order to understand the discrepancy.
\end{abstract}

\section{Introduction}

The collective mode properties of the ground state within  the FQHE 
research activity has been the subject of interest in the literature 
~\cite{1,3,10,2}.
A pioneering paper on this theme \cite{1}
is based on some ideas already introduced by Feynman \cite{fey}
in the theory of Superfluidity.
The general conclusion of this work indicated the 
stability of the considered ground
 state due to the appearance of a gap in the
spectrum of the lowest energy collective excitation. 

More recently, in ~\cite{3}, the subject has been  considered through
a Chern-Simons formulation of the composite fermion model. 
However, these results 
only cover the small wavevector  range and cannot
furnish appreciable information 
on the ground state stability properties. 

In a previous paper~\cite{2} we studied  the collective mode
excitation for composite fermions at 1/3 filling factor.
In that  work the Bethe-Salpeter equation was considered and through
it  the dispersion relation was calculated.
The result obtained was unexpected since  instead of
being consistent with  to the one obtained by Girvin 
et. al. \cite{1}, we observed an unstable behavior. 

Seemingly this result would be a consequence of the known 
perturbation theory difficulties of the composite fermions model.
On the other hand such conclusion could be compatible with recent
argument about the posible  role of cristalline properties
in the ground state \cite{11,12,13,14}.

In order to get a better understanding of the problem, 
in this note we want to explore
what would be the result  in the  case
of considering phenomenologically the two dimensional electron 
gas in a
Laughlin state at 1/3 filling factor. The technique to be used was 
similar  to the one  employed in the previous work Ref.~\cite{2}.

The conclusion coincided with the one  in Ref~\cite{2}. It was 
detected an instability which appears for a wide range of values
of the gap parameter $\Delta$ describing
phenomenologically the  one-particle gap in the Laughlin state
propagator (~\cite{2}). 

It is better emphasized that ~\cite{1}
leads to conclusions
matching satisfactorily with the spectrum   
calculated numerically starting from the study of  systems of small
number of particles \cite{hal}, and with the experimental results
\cite{exp}.
However, the determination or not  for  the presence of crystalline
properties in the ground state can be expected to require higher
number of particles than  the ones considered up to now. Moreover, 
the assumptions employed in ~\cite{1} in order to conclude the
ground state stability are physically argumented and motivating,
but not by far unquestionable, that is why the convenience remains
yet for further investigations on the collective mode properties.

We divide the present letter in four sections.
As a matter of checking the procedures, 
the second section is devoted to  reproduce the dispersion
relation in the already known situation 
when the first Landau is fully occupied. The result obtained
correctly resembles  the spectrum obtained  in Ref.~\cite{4}.

In the second part  the fractional Landau level occupied at 1/3
filling factor is discussed after  using a suitable dependence on
the  frequency and coordinates for the one particle propagator.
This construction was  based on the  general results and arguments  
 given in Refs.~\cite{5,gir,7}.
Employing  the  magnetoroton basis it follows the diagonalization
of the BSE in the approximation of retaining only the first Landau
level states. 
 
In the fourth  section the inclusion of the first excited Landau
level in the considerations is done. The outcome indicates that the
collective mode frequency approaches the cyclotronic resonance at
zero momentum as in Ref.~\cite{2}.
Therefore, the  obtained results reproduce the general properties of 
the collective mode following from the composite fermion BSE 
investigation in our previous paper~\cite{2}.
Finally, in the last section possible extensions of the work and 
connections with related research are commented.

\section{Collective mode for one filled  Landau level }

The BSE is considered here in order to reproduce  
the known  collective excitation results  in the situation 
when only  one Landau level is completely filled. The aim is to check
the results of  the technical procedures to be used in the following 
sections in this solved problem. 
The discussion closely follows the one in Ref~\cite{2} the difference
being related with the absence of anyon like interactions.
 In such  case the equation has the form

\[
{\cal F}(1,1,2,2')={\cal F}_{o}(1,1',2,2')
\]

\begin{equation}
+\int\int\int d4\,d5\,d6{\cal F}_{o}(1,1',3,4){\cal W}(3,4,5,6)
{\cal F}(5,6,2',2), 
\label{1}
\end{equation}

\noindent
where ${\cal F}_o$ is the interaction free four-point function and
 the  interaction kernel in the first approximation  is taken as
the following  functional derivative

\[
{\cal W}={\it i}\frac{\delta {\cal W}_{HF}(3,4)}{\delta {\it
G}(5,6)} 
\]

\noindent
where ${\cal W}_{HF}$ is the mass operator in the Hartree-Fock
approximation.

 It is possible to simplify the integral
equation (\ref{1}) after introducing magneto-exciton 
wave functions Ref.~\cite{8,9} which have the explicit form

\begin{eqnarray}
\Psi_{n\alpha}^{n'}&=&\frac{(-1)^n}{L}\frac{1}
{(2\pi 2^{n+n'}n!n'!)^{1/2}}
\left(2\frac{\partial}{\partial
z_{1}^*}-\frac{1}{2}z_{1}\right) ^n\left(2\frac{\partial}
{\partial z_{2}}-\frac{1}{2}z_{2}^*\right)^{n'}
\nonumber \\
& & \exp[{-(|z_{1}|^2+|z_{2}|^2+|z_{\alpha}|^2)/4}] 
\exp[{(z_{1}^*z_{2}+z_{1}^*z_{\alpha}-z_{2}z_{\alpha}^*)/2}], 
\\ z&=&x-y i .
\end{eqnarray}

\noindent
where $L$ is the radius of the macroscopic sample.
These two-particle states  are  constructed from the one-body wave
functions for electrons and holes in the
 $n$th and $n^{\prime}$th Landau levels.
They are characterized by a center of mass momentum which can be
written in complex form as $z_{\alpha}={\it i} q_{x} -q_{y}$.
The main simplification introduced by these states is that the
$z_{\alpha}$ is conserved and all
the matrix elements are diagonal in this quantum number
Ref.~\cite{8}. 

Then, through the use the basis functions (2), 
the integral  equation (\ref{1}) can be written  as a linear matrix
equation in the form

\[
{\cal F}_{nm,n',m'}(\alpha,\omega)=\delta_{nn'}\delta_{mm'}
{\cal F}_{o\,nm}(\omega)
\]
\begin{equation}
+\sum_{kl}{\cal F}_{o\,nm}(\omega)<nm\alpha|{\cal W}|kl\alpha>{\cal 
F}_{kl,n'm'}(\alpha,\omega).
\label{2}
\end{equation}

The matrix elements of the free four point function ${\cal F}_o$ and
the interaction kernel ${\cal W}$ are given by the expressions

\begin{eqnarray}
\left\langle
\begin{array}{ll}
m'\\
m   
\end{array}
\left|{\cal F} \right|
\begin{array}{ll}
n'\\
n   
\end{array}
\right\rangle 
=\int\int\int\int\Psi_{m\alpha }^{*m'}(\vec{1},\vec{4}){\cal
F}_{o}(\vec{1},\vec{4}|\vec{2},\vec{3})\Psi_{n\alpha}^{n'}
(\vec{2},\vec{3})
d\vec{1}d\vec{2}d\vec{3}d\vec{4},
\end{eqnarray} 

\begin{eqnarray}
\left\langle
\begin{array}{ll}
m'\\
m
\end{array}
|{\cal W}|
\begin{array}{ll}
n'\\
n
\end{array}
\right\rangle
=\int\int\int\int\Psi_{m\alpha}^{*m'}(\vec{1},\vec{4}){\cal 
W}(\vec{1},\vec{4}|\vec{2},\vec{3})\Psi_{n\alpha}^{n'}
(\vec{2},\vec{3})
d\vec{1}d\vec{2}d\vec{3}d\vec{4}, 
\end{eqnarray}

\noindent
in which the dependence on the $\alpha$ quantum number has been omitted
and a vertical representation of the bracket indexes is also used.

The interaction kernel in the here considered first approximation
consists in the sum
of two contributions coming from the Coulomb interaction
and mainly corresponds to the  disregarding of  the screening corrections

\[
{\cal W}={\cal W}^{C}_{12}+{\cal W}^{C}_{22}.
\]

\noindent
At experimental FQHE regime  this approximation seems to be reasonable
because at magnetic length distances the screening of the coulomb 
interaction should  be weak since  the mean distance between electrons
is of the same order than the magnetic lenght in experimental samples.
The explicit formulae for the ${\cal W}^{C}$ terms
 and for ${\cal F}_{o}$ are given in the Appendix.

The collective modes of the system, corresponds with the lowest
frequencies leading to zero eigenvalues of the inverse kernel

\begin{equation}
{\cal F}^{-1}=({\cal F}_{o})^{-1} - {\cal W},
\end{equation}

\noindent
which in addition can produce singularities of the dielectric response
tensor temporal-temporal component~\cite{8}.
Due to the fact that ${\cal F}_{o}$ is qualitatively different for
particle-hole and hole-particle channels it becomes useful to divide
the matrix equation~(\ref{2}) in four sub-blocks.
Basically, the block representation  is obtained by restricting,  
as conceived  in Ref.~\cite{8}, the indexes $m$ and $n$ to empty Landau
levels and $m'$ and $n'$ 
to filled ones and define the operators  ${\cal E}$ and ${d}$
through their block matrix elements 

\[
\left\langle\begin{array}{ll}
m'\\
m
\end{array}
\left|{\cal W}\right|
\begin{array}{ll}
n'\\
n
\end{array}
\right\rangle  \equiv
\left\langle\begin{array}{ll}
m'\\
m
\end{array}
\left|{\cal E}\right|
\begin{array}{ll}
n'\\
n
\end{array}
\right\rangle,
\]

and

\[
\left\langle\begin{array}{ll}
m'\\
m
\end{array}
\left|{\cal W}\right|
\begin{array}{ll}
n\\
n'
\end{array}
\right\rangle  \equiv
\left\langle\begin{array}{ll}
m'\\
m
\end{array}
\left| d\right|
\begin{array}{ll}
n'\\
n
\end{array}
\right\rangle,
\]
\noindent
for which the properties of the magneto-exciton wave function also
implies

\[
\left\langle\begin{array}{ll}
m\\
m'
\end{array}
\left|{\cal W}\right|
\begin{array}{ll}
n\\
n'
\end{array}
\right\rangle  \equiv
\left\langle\begin{array}{ll}
m'\\
m
\end{array}
\left|{\cal E}\right|
\begin{array}{ll}
n'\\
n
\end{array}
\right\rangle^*,
\]

\[
\left\langle\begin{array}{ll}
m\\
m'
\end{array}
\left|{\cal W}\right|
\begin{array}{ll}
n'\\
n
\end{array}
\right\rangle  \equiv
\left\langle\begin{array}{ll}
m'\\
m
\end{array}
\left|d\right|
\begin{array}{ll}
n'\\
n
\end{array}
\right\rangle^*.
\]

\noindent
The explicit expressions for the matrix elements of the operators 
${\cal E}$  and $d$ are given in the Appendix.

\noindent
Moreover, again  introducing  an operator $\Delta\epsilon$ through
its matrix elements  as 

\begin{equation}
\left\langle
\begin{array}{ll}
m'\\
m
\end{array}
\left|\Delta\epsilon\right|
\begin{array}{ll}
n'\\
n
\end{array}
\right\rangle
\equiv\delta_{mn}\delta_{m'n'}(\epsilon_{n}-\epsilon_{n'})
\end{equation}

\noindent
in order to compact the blocks coming from the matrix representation
of ${\cal F}_{o}$, the  Bethe-Salpeter equation (4)
can be reduced to the matrix form

\[
\left\langle
\begin{array}{ll}
m'\\
m
\end{array}
\left|{\cal F}\right|
\begin{array}{ll}
n'\\
n
\end{array}
\right\rangle
\equiv \int\int\int\Psi_{m\alpha}^{m'*}(r_{1},r_{4}){\cal 
F}(r_{1},r_{4}|r_{2}r_{3})\Psi_{n\alpha}^{n'}(r_{2},r_{3})
dr_{1}dr_{2}dr_{3}dr_{4} \]

\begin{eqnarray}
= \left\langle
\begin{array}{ll}
m'\\
m
\end{array}
\left|
\left[\begin{array}{cc}
\omega-\Delta\epsilon-{\cal E}+{\it i}\eta & -d\\ 
-d^{+} & -\omega-\Delta\epsilon-{\cal E}^{*}+{\it i}\eta
\end{array}
\right]^{-1}
\right|
\begin{array}{ll}
n'\\
n
\end{array}
\right\rangle.\label{x}
\end{eqnarray}

This representation was then used to determine numerically the
collective mode dispersion relation. For this purpose a finite
number of basis functions corresponding to the Landau levels
index up to a maximum value were retained in constructing the
BS matrix (4).
The roots of the determinant of the matrix (\ref{x}) were found.
The spectrum of the collective mode is shown in Fig.1.
It can be seen that the behavior is qualitatively similar to
the one obtained in Ref.~\cite{4}.
In the next section the  framework for the discussion of the
collective mode is somewhat modified intending
 to describe the $1/3$ partially
filled Landau level.

\section{Partially filled Landau level}

Let us  analyze now the situation  when one Landau level is partially
occupied at 1/3 filling factor. After projecting BSE on the first
Landau level is possible diagonalize it by using the magnetoexciton
basis states  with $n=n'=0$ in the following way

\begin{equation}
{\cal F}_{00,00}(\alpha,w)={\cal F}_{00}(w) + \sum {\cal
F}_{00}(\omega)<\alpha,0,0|{\cal W}|\alpha,0,0>
{\cal F}_{00,00}(\alpha, w)
\label{e}
\end{equation}

\noindent
Note that all ${\cal F}_{00,00}$, ${\cal F}_{00}$ and ${\cal W}$ have
been projected in the same subspace of the $n=n'=0$ functions.   
The one-particle propagator will be also projected in the first Landau
level and its spatial structure can be explicitly written after
considering the results of Refs.~\cite{5,gir,7}.
The concrete expression to be used has the form

\[
G(\vec{x},\vec{x}',w)=\Pi(\vec{x},\vec{x}')G^e(w),
\]

\noindent
where $G^e(w)$ will be defined in a phenomenological fashion in order
to model the exact structure associated to the Laughlin's state, and
$\Pi$ is the projector
operator in the first Landau level (See Appendix).
Concretely, it  will be  assumed that the frequency dependence is 
given by two poles corresponding to energies shifted in a
one-particle gap as follows

\[
G^e(w)=\left[\frac{1}{3}\frac{1}{w+\Delta/2+i\eta}+\frac{2}{3}
\frac{1}{w-\Delta/2-i\eta}\right]
\]

\noindent
where the frequency and energies in this section are measured in units
of $e^2/r_{o}$ and the factors $1/3$ and $2/3$ assures the form of the
exact one-particle density matrix for the Laughlin state determined
in Refs.~\cite{5,7}.
The one-particle energy reference system is taken at midway between
the two energies $+\Delta/2$ and $-\Delta/2$.
It turns out that this dependence seems to be a good description of
the exact one according to the arguments given by Rezayi in
Ref.~\cite{7}.

After fixed the  $G^e(w)$ dependence, the quantity   ${\cal F}_{oo}$ 
in  (\ref{e}) can be calculated explicitly and takes the form

\begin{equation}
{\cal F}_{o}=\frac{2}{9}\left[\frac{1}{w-\Delta-i\eta}
+ \frac{1}{-w-\Delta-i\eta}
\right].
\end{equation}

\noindent
The numerical value for gap parameter  $\Delta$ can be
 estimated from the results in
Ref.~\cite{7}.
The homogeneous BSE associated to Eq~(\ref{e}) now reduces to a simple
scalar condition 

\begin{equation} 
{\cal F}_{o}^{-1}-<\alpha,0,0|{\cal W}|\alpha,0,0>=0. 
\label{ga}
\end{equation}

\noindent
Relation (12) can be simplified to write the explicit 
formula for the dispersion relation

\begin{eqnarray}
w^2-\Delta ^{2} &=&-\frac{4}{9}\,\Delta.\left [(\pi/2)^{1/2}
e^{-\vec{q}^2/4}
{\cal I}_{o}(\vec{q}^2/4)-\frac{1}{\mid\vec{q}\mid }
e^{-\vec{q}^2/2}\right]\label{a},\\
\vec{q}^2&=&q_x^2+q_y^2  \nonumber
\end{eqnarray}

\noindent
where the term in brackets corresponds with the evaluation of
${\cal W}_{c}$ when $n=n'=0$ from the expressions in the Appendix
after setting $u_c=1$.
  
The spectrum ~(\ref{a}) predicts  instability  for a  range of
the phenomenological one-particle  gap parameter $\Delta$. 
The Fig.2 shows the momentum dependence  of the collective mode
frequency for the estimated
gap parameter value  $\Delta\equiv 0.2$  measured in units of
$~e^2/(\epsilon l_{o})$ as taken from Ref.~\cite{7}.

It should be mentioned that 
Giulianni and Quinn ~\cite{10} had also pointed out a similar
behavior through an study of the BSE associated to  the Tao-Thouless
parent states (TT).
However, in that system  the instability result was attributed to the
ground state degeneration of the TT state. The present description,
however, seems to be well designed for 
the consideration of the collective mode 
in the Laughlin's or any translationally invariant temptative ground
state.

The instability region for the parameter $\Delta$ is given 
by

\[
\Delta  < 0.2474.
\]

\noindent

Therefore for an appreciable  range of values for the parameter  
$\Delta$ describing the one-particle gap,
 an instability under the formation of periodic
charge density waves results.  

\section{Inclusion of one higher Landau level}

In Ref.~\cite{2} the collective mode dispersion relation at low
momenta satisfied the Kohn theorem in an external magnetic field.
In other words the low momenta frequency approached the cyclotron frequency.
In the previous section, where only the lowest Landau level was
considered it should  not be possible to test this property.
However the validity of such a condition
is important because it is  related to the general 
result stating that the
zero momentum cyclotron resonance saturates the oscillator
strength sum rule once the  homogeneity of the system is assumed.
It can be noticed  that the collective mode spectrum 
evaluated in  Ref.~\cite{2} at
low momentum seems to be compatible with such property.

Then, the purpose of this section is to consider corrections
for the  low momentum behavior of the  collective mode,
introduced by the inclusion of the next empty Landau level.
 In this way it will be intended  to check for the validity of  Kohn
 theorem in the present approach.

Our procedure will be  an slight modification of the one
used in the first section. The form of matrix element of
${\cal F}_{o}$ will be taken in the approximation consisting in
retaining only the lowest two Landau levels n=0,n=1. This
approximation produces for the relevant matrix elements of
${\cal F}_0$ among the retained states the expression

\medskip
\medskip

\[
\left\langle
\begin{array}{ll}
m'\\
m
\end{array}
\left|{\cal F}_o \right|
\begin{array}{ll}
n'\\
n
\end{array}   
\right\rangle
= \delta_{\alpha \beta} \delta_{mn} \delta_{m' n'}
.\left[
\begin{array}{lll}
(\omega-(\epsilon_{n}-\epsilon_{n'})+i \eta)^{-}1&\small{n=0,\,
 n'=1}\\
(-\omega-(\epsilon_{n'}-\epsilon_{n})+i\eta)^{-1}&
\small{ n'=0,\,\,n=1} \\
\frac{2}{9}((w-\Delta'-i\eta)^{-1}+(-w-\Delta'-i\eta)^{-1})&
\small{n=0,\,n'=0,}\\
0  &\small{n=1,\,n'=1,}
\end{array}\right.
\label{sx}
\]

\noindent
corresponding to  neglecting the higher $n>1$ Landau levels.
A new parameter  $\Delta'= u_c\Delta$ has been introduced 
because the energy in this section is being measured in units of
$\hbar w_c$ .

\noindent 
The matrix elements of ${\cal W}$ have now mixing of contributions
${\cal E}_{00}$, ${\cal E}_{01}$, ${\cal E}_{10}$, $d_{01}$,
$d_{10}$. Note that due to

  $$\left\langle 1,1 \mid{\cal F}_0\mid 1,1 \right\rangle=0$$

\noindent
and ${\cal F}_0$ being  diagonal in the considered  particle-hole
basis, imply that all the matrix elements of ${\cal W}$ among the
state n=1, m=1 and any of the other three  retained states decouple
from equation (4).

Then, the  dispersion relation condition for the collective
mode following from the homogeneous BSE
takes now the form 

\begin{eqnarray}
Det
\left[ 
\begin{array}{ccc}
\frac{9}{4}\frac{(w^2-\Delta'^2)}{\Delta'}-{\cal E}_{00}&- d_{01}&
-{\cal E}_{01}\\
-d_{10}^{*}&\omega-\Delta\epsilon-{\cal E}_{11}+{\it i}\eta &
-d_{11}\\ 
-{\cal E}_{01}^{*}&-d_{11}^{*} & -\omega-\Delta\epsilon-
{\cal E}_{11}^{*}
+{\it i}\eta
\end{array}
\right] = 0\nonumber\\
\label{x1}
\end{eqnarray}

\noindent
The numerical solution of (\ref{x1}) for the same value
 $\Delta=0.2$ chosen in Section 3
produces the collective mode dispersion
relation shown in Fig.3 where  the instability region is again
present.
However, at low momentum it follows that the frequency 
of the lowest energy collective mode instead of tending
to infinity as in Fig.2, reaches the cyclotron resonance. 
Therefore,  the present  discussion reproduces
 in this limit the conclusion of
Ref.~\cite{2} and  the lowest energy 
collective mode again  satisfies the Kohn Theorem. 
In a similar way as in Ref. ~\cite{3} there are two modes that
(in the present  case approximately) approach to the cyclotron
resonance
in the low momentum limit. In Fig.4, a magnified part of the low
momentum region of Fig. 3, shows how  the inter-Landau level
collective mode at low momentum gets the divergent frequency behavior
which the intra-Landau mode (Fig.1)  previously had
in the preceding approximation. 
Observe that in Section 3, as well as here, this divergent behavior
at low momentum seems to be forced  with the saturation of the
oscillator strength sum rule by the cyclotronic resonance.

\section{Comments}

We have obtained indications of the presence of unstable  excitations
in a translationally invariant states for FQHE systems corresponding
to the developing of crystalline charge density waves. Such a result
is coherent with the search in progress 
for ground states showing periodic charge density oscillations 
~\cite{11,12,13,14}. The wavefunctions discussed in these 
works  can  exhibit  relatively weak 
periodic charge densities in a lattice having one flux quantum
passing through its unit cell. The associated wave vector
of this periodicity has similar magnitude as the momentum values
 laying inside the  instability region discussed here. In fact the 
previous consideration of these states motivated
 our interest in discussing
the stability properties  of 
the homogeneous ground states through the
investigation of  their collective modes.
  
It is clear that  further research about
the discrepancy between the results 
of this work with the conclusions following from the single mode
approximation in Ref.~\cite{1} is in need. In this direction 
some speculative reasoning which we consider useful to expose  will be
given below.

 Under assuming the existence of a sufficiently weak
 charge density wave in the true ground state of the system,
 the correlation functions 
of this wavefunction  would be well approximated by the ones associated
to the Laughlin
state. Therefore, the single mode approximation for the lowest energy 
excitations  
of this real ground state (determined by such correlations)  
would be expected to produce similar results to the ones in
Ref.~\cite{1}.
However, in another hand, if an exactly translationally invariant
candidate for ground state (like the Laughlin's one)  is unstable,
the single mode approximation should  not be valid. But, in such
conditions the ansatz wavefunction given by the Fourier components of
the density would also become a sort of imposition eventually capable
of ruling out the instability result from the outcome in Ref.
~\cite{1}.
Henceforth, if this considerations have some foundation the
similarity of the results for the single mode approximation for
collective mode with the essentially exact data for the energy
spectrum for few particle systems would become a possible  outcome
in spite of presence of a charge density wave instability of  the
considered state.

The investigation on these questions will be the objective of a
future continuation of the work.

\section{Acknowledgments.} 

It is pleasure to express our gratitude to our colleague Augusto
Gonzalez for important discussions and we also appreciate the
suggestions made by Gerardo Gonzalez and Jesus Sanchez.
 We all are also deeply indebted  to the Third World Academy of
Sciences for its support through  the TWAS Research Grant 93-120
RG/PHYS/LA.

\newpage

\begin{center}
{\bf APPENDIX}
\end{center}

\subsection{{\bf Definitions of various auxiliary quantities}}

\[
c(m,n)=(-1)^{m + n}\frac{z_{\alpha}^n z_{\alpha}^{*m}}
 {(2^{m+n}m!n!)^{1/2}},
\]

\[
d(m,n)=(-1)^{m + n}\frac{z_{\alpha}^{*(m+n)}}{(2^{m+n}m!n!)^{1/2}},
\]

\[
co(m,n)=\frac{m!}{(m-l)!l!},
\]

\[
b=\frac{1}{2}|z_{\alpha}|^2
\]

The  projection operator in the first Landau level $\Pi$ is
defined as

\[
\Pi(1,2)={1\over 2 \pi l_0^2 } \exp {[-(|z_1|^2+|z_2|^2)/4+z_{1}^{*}
 z_2/2}].
\]
\noindent
with $l_{o}$ is the magnetic length in the external magnetic field
$B$.

\subsection{{\bf The selfenergies and coulomb contributions for each
Landau level of index ${n}$}}

\[
{\epsilon}_{n}=n +{\epsilon}_{c}(n)
\]

\[
{\epsilon}_{c}(n)= - u_c (\pi/2)^{1/2}\left(-1 + \sum_{l=0}^n
\frac{(-1/2)^ln!
(2l-1)!!)}{l!^2(n-l)!}\right)
\]

\noindent
where $n$ is the index of the Landau level in the magnetic field and
the constant $u_c$ is given by

\[
u_c=\frac{e^2}{\epsilon l_o}\left/\frac{\hbar e B}{m c}\right.
\]

\subsection{{\bf The matrix elements of ${\cal F}_o$  for fully empty
or filled Landau levels}}

\[
\left\langle
\begin{array}{ll}
m'\\
m
\end{array}
\left|{\cal F}_o \right|
\begin{array}{ll}
n'\\
n
\end{array}   
\right\rangle
=\delta_{\alpha \beta} \delta_{mn} \delta_{m' n'}
\left[
\begin{array}{cc}
(\omega-(\epsilon_{n}-\epsilon_{n'})+i \eta)^{-1} &
\mbox{${n}$\,\small{empty},\, ${n'}$\,\,\small{occupied}} \\
(-\omega-(\epsilon_{n'}-\epsilon_{n})+i\eta)^{-1} & 
\mbox{${n'}$\small{empty,}\,\, ${n}$\,\,\small{occupied}} \\
0 &\mbox{\small{otherwise}} \\
\end{array}
\right.
\]

\noindent
when the first Landau level is not filled a nonzero value for the\\  
matrix elements $<0,0|{\cal F}_{o}|0,0>$ can appear.

\subsection{{\bf Matrix elements of interaction kernel ${\cal W}$}}

\[
{\cal W}_{21}^{C}=-\int d1\int d2 {u_c \over |1-2|}\Psi_{A}^*(1,2)
\Psi_{B}(1,2),
\]

\[
{\cal W}_{22}^{C}=\int d1\int d2\int d3 {u_c \over |1-2|}
 \Psi_{A}^*(2,2)\Psi_{B}(1,1).
\]

\subsection{{\bf The ${\cal E}$ block matrix components for all $m$,
$n$}}

\[
{\cal E}_{21}(m,n)=
- u_c \,c(m,n)e^{-b}\sum_{l=0}^{m}\sum_{k=0}^n
co(m,l)co(n,k)\Gamma(\frac{(1+k+l+|l-k|)}{2})
\]

\[ 
.(1/2)^{|l-k|/2-(l+k)/2}(-1)^{-(l+k)}|z_{\alpha}|^{|l-k|-(l+k)}\]
\[
.\frac{{}_{1}F_{1}((1+k+l+|l-k|)/2,1+|l-k|,\frac{|z_{\alpha}|^2}{2})}
{{2}^{1/2}\Gamma(1+|l-k|)},
\]

\[
{\cal E}_{22}(m,n)=u_c\, c(m,n)\frac{e^{-b}}{|z_{\alpha}|},
\]

\[
d_{21}(m,n)=-u_c\, d(m,n)e^{-b}\sum_{l=0}^{m+n}co(m,l)(1/4)^{|l|/2}
(-1)^{-l}|z_{\alpha}|^{|l|-l}\Gamma(|l|+1/2)
\]

\[ 
.\frac{{}_{1} F_{1}(|l|+1/2,|l|+1,\frac{|z_{\alpha}|^{2}}{2})}
{\sqrt{2}(1/2)^{|l|}\Gamma(|l|+1)},
\]

\[
d_{22}(m,n)=u_c \, d(m,n)\frac{e^{-b}}{|z_{\alpha}|}.
\]


\noindent
\figure{The collective mode energy for the one filled Landau
level\label{Fig.1}}

\figure{The collective mode energy for 1/3 filled Landau level as
 disregarding the higher ones\label {Fig.2}}
\noindent
\figure{The low momentum behavior of the collective mode for 1/3 filled 
Landau level including the first excited level\label{Fig.3}}
\noindent
\figure{The inter and intra Landau level collective mode energy at 1/3 
filled Landau level\label{Fig.4}}

\newpage
\setlength{\unitlength}{0.240900pt}
\ifx\plotpoint\undefined\newsavebox{\plotpoint}\fi
\sbox{\plotpoint}{\rule[-0.500pt]{1.000pt}{1.000pt}}%
\begin{picture}(1500,900)(0,0)
\font\gnuplot=cmr10 at 10pt
\gnuplot
\sbox{\plotpoint}{\rule[-0.500pt]{1.000pt}{1.000pt}}%
\put(220.0,113.0){\rule[-0.500pt]{1.000pt}{173.207pt}}
\put(220.0,113.0){\rule[-0.500pt]{4.818pt}{1.000pt}}
\put(198,113){\makebox(0,0)[r]{1.05}}
\put(1416.0,113.0){\rule[-0.500pt]{4.818pt}{1.000pt}}
\put(220.0,193.0){\rule[-0.500pt]{4.818pt}{1.000pt}}
\put(198,193){\makebox(0,0)[r]{1.1}}
\put(1416.0,193.0){\rule[-0.500pt]{4.818pt}{1.000pt}}
\put(220.0,273.0){\rule[-0.500pt]{4.818pt}{1.000pt}}
\put(198,273){\makebox(0,0)[r]{1.15}}
\put(1416.0,273.0){\rule[-0.500pt]{4.818pt}{1.000pt}}
\put(220.0,353.0){\rule[-0.500pt]{4.818pt}{1.000pt}}
\put(198,353){\makebox(0,0)[r]{1.2}}
\put(1416.0,353.0){\rule[-0.500pt]{4.818pt}{1.000pt}}
\put(220.0,433.0){\rule[-0.500pt]{4.818pt}{1.000pt}}
\put(198,433){\makebox(0,0)[r]{1.25}}
\put(1416.0,433.0){\rule[-0.500pt]{4.818pt}{1.000pt}}
\put(220.0,512.0){\rule[-0.500pt]{4.818pt}{1.000pt}}
\put(198,512){\makebox(0,0)[r]{1.3}}
\put(1416.0,512.0){\rule[-0.500pt]{4.818pt}{1.000pt}}
\put(220.0,592.0){\rule[-0.500pt]{4.818pt}{1.000pt}}
\put(198,592){\makebox(0,0)[r]{1.35}}
\put(1416.0,592.0){\rule[-0.500pt]{4.818pt}{1.000pt}}
\put(220.0,672.0){\rule[-0.500pt]{4.818pt}{1.000pt}}
\put(198,672){\makebox(0,0)[r]{1.4}}
\put(1416.0,672.0){\rule[-0.500pt]{4.818pt}{1.000pt}}
\put(220.0,752.0){\rule[-0.500pt]{4.818pt}{1.000pt}}
\put(198,752){\makebox(0,0)[r]{1.45}}
\put(1416.0,752.0){\rule[-0.500pt]{4.818pt}{1.000pt}}
\put(220.0,832.0){\rule[-0.500pt]{4.818pt}{1.000pt}}
\put(198,832){\makebox(0,0)[r]{1.5}}
\put(1416.0,832.0){\rule[-0.500pt]{4.818pt}{1.000pt}}
\put(220.0,113.0){\rule[-0.500pt]{1.000pt}{4.818pt}}
\put(220,68){\makebox(0,0){0}}
\put(220.0,812.0){\rule[-0.500pt]{1.000pt}{4.818pt}}
\put(423.0,113.0){\rule[-0.500pt]{1.000pt}{4.818pt}}
\put(423,68){\makebox(0,0){1}}
\put(423.0,812.0){\rule[-0.500pt]{1.000pt}{4.818pt}}
\put(625.0,113.0){\rule[-0.500pt]{1.000pt}{4.818pt}}
\put(625,68){\makebox(0,0){2}}
\put(625.0,812.0){\rule[-0.500pt]{1.000pt}{4.818pt}}
\put(828.0,113.0){\rule[-0.500pt]{1.000pt}{4.818pt}}
\put(828,68){\makebox(0,0){3}}
\put(828.0,812.0){\rule[-0.500pt]{1.000pt}{4.818pt}}
\put(1031.0,113.0){\rule[-0.500pt]{1.000pt}{4.818pt}}
\put(1031,68){\makebox(0,0){4}}
\put(1031.0,812.0){\rule[-0.500pt]{1.000pt}{4.818pt}}
\put(1233.0,113.0){\rule[-0.500pt]{1.000pt}{4.818pt}}
\put(1233,68){\makebox(0,0){5}}
\put(1233.0,812.0){\rule[-0.500pt]{1.000pt}{4.818pt}}
\put(1436.0,113.0){\rule[-0.500pt]{1.000pt}{4.818pt}}
\put(1436,68){\makebox(0,0){6}}
\put(1436.0,812.0){\rule[-0.500pt]{1.000pt}{4.818pt}}
\put(220.0,113.0){\rule[-0.500pt]{292.934pt}{1.000pt}}
\put(1436.0,113.0){\rule[-0.500pt]{1.000pt}{173.207pt}}
\put(220.0,832.0){\rule[-0.500pt]{292.934pt}{1.000pt}}
\put(45,472){\makebox(0,0){$\omega/\omega_{c}$}}
\put(828,23){\makebox(0,0){$ql_{o}$}}
\put(1356,742){\makebox(0,0){Fig.1}}
\put(220.0,113.0){\rule[-0.500pt]{1.000pt}{173.207pt}}
\put(240,114){\raisebox{-.8pt}{\makebox(0,0){$\Diamond$}}}
\put(321,329){\raisebox{-.8pt}{\makebox(0,0){$\Diamond$}}}
\put(342,357){\raisebox{-.8pt}{\makebox(0,0){$\Diamond$}}}
\put(382,385){\raisebox{-.8pt}{\makebox(0,0){$\Diamond$}}}
\put(423,382){\raisebox{-.8pt}{\makebox(0,0){$\Diamond$}}}
\put(463,359){\raisebox{-.8pt}{\makebox(0,0){$\Diamond$}}}
\put(504,324){\raisebox{-.8pt}{\makebox(0,0){$\Diamond$}}}
\put(524,307){\raisebox{-.8pt}{\makebox(0,0){$\Diamond$}}}
\put(625,254){\raisebox{-.8pt}{\makebox(0,0){$\Diamond$}}}
\put(828,425){\raisebox{-.8pt}{\makebox(0,0){$\Diamond$}}}
\put(1031,634){\raisebox{-.8pt}{\makebox(0,0){$\Diamond$}}}
\put(1233,745){\raisebox{-.8pt}{\makebox(0,0){$\Diamond$}}}
\put(1436,812){\raisebox{-.8pt}{\makebox(0,0){$\Diamond$}}}
\end{picture}

\vspace{2cm}

\setlength{\unitlength}{0.240900pt}
\ifx\plotpoint\undefined\newsavebox{\plotpoint}\fi
\sbox{\plotpoint}{\rule[-0.500pt]{1.000pt}{1.000pt}}%
\begin{picture}(1500,900)(0,0)
\font\gnuplot=cmr10 at 10pt
\gnuplot
\sbox{\plotpoint}{\rule[-0.500pt]{1.000pt}{1.000pt}}%
\put(220.0,113.0){\rule[-0.500pt]{292.934pt}{1.000pt}}
\put(220.0,113.0){\rule[-0.500pt]{1.000pt}{173.207pt}}
\put(220.0,113.0){\rule[-0.500pt]{4.818pt}{1.000pt}}
\put(198,113){\makebox(0,0)[r]{0}}
\put(1416.0,113.0){\rule[-0.500pt]{4.818pt}{1.000pt}}
\put(220.0,185.0){\rule[-0.500pt]{4.818pt}{1.000pt}}
\put(198,185){\makebox(0,0)[r]{0.1}}
\put(1416.0,185.0){\rule[-0.500pt]{4.818pt}{1.000pt}}
\put(220.0,257.0){\rule[-0.500pt]{4.818pt}{1.000pt}}
\put(198,257){\makebox(0,0)[r]{0.2}}
\put(1416.0,257.0){\rule[-0.500pt]{4.818pt}{1.000pt}}
\put(220.0,329.0){\rule[-0.500pt]{4.818pt}{1.000pt}}
\put(198,329){\makebox(0,0)[r]{0.3}}
\put(1416.0,329.0){\rule[-0.500pt]{4.818pt}{1.000pt}}
\put(220.0,401.0){\rule[-0.500pt]{4.818pt}{1.000pt}}
\put(198,401){\makebox(0,0)[r]{0.4}}
\put(1416.0,401.0){\rule[-0.500pt]{4.818pt}{1.000pt}}
\put(220.0,473.0){\rule[-0.500pt]{4.818pt}{1.000pt}}
\put(198,473){\makebox(0,0)[r]{0.5}}
\put(1416.0,473.0){\rule[-0.500pt]{4.818pt}{1.000pt}}
\put(220.0,544.0){\rule[-0.500pt]{4.818pt}{1.000pt}}
\put(198,544){\makebox(0,0)[r]{0.6}}
\put(1416.0,544.0){\rule[-0.500pt]{4.818pt}{1.000pt}}
\put(220.0,616.0){\rule[-0.500pt]{4.818pt}{1.000pt}}
\put(198,616){\makebox(0,0)[r]{0.7}}
\put(1416.0,616.0){\rule[-0.500pt]{4.818pt}{1.000pt}}
\put(220.0,688.0){\rule[-0.500pt]{4.818pt}{1.000pt}}
\put(198,688){\makebox(0,0)[r]{0.8}}
\put(1416.0,688.0){\rule[-0.500pt]{4.818pt}{1.000pt}}
\put(220.0,760.0){\rule[-0.500pt]{4.818pt}{1.000pt}}
\put(198,760){\makebox(0,0)[r]{0.9}}
\put(1416.0,760.0){\rule[-0.500pt]{4.818pt}{1.000pt}}
\put(220.0,832.0){\rule[-0.500pt]{4.818pt}{1.000pt}}
\put(198,832){\makebox(0,0)[r]{1}}
\put(1416.0,832.0){\rule[-0.500pt]{4.818pt}{1.000pt}}
\put(220.0,113.0){\rule[-0.500pt]{1.000pt}{4.818pt}}
\put(220,68){\makebox(0,0){0}}
\put(220.0,812.0){\rule[-0.500pt]{1.000pt}{4.818pt}}
\put(372.0,113.0){\rule[-0.500pt]{1.000pt}{4.818pt}}
\put(372,68){\makebox(0,0){1}}
\put(372.0,812.0){\rule[-0.500pt]{1.000pt}{4.818pt}}
\put(524.0,113.0){\rule[-0.500pt]{1.000pt}{4.818pt}}
\put(524,68){\makebox(0,0){2}}
\put(524.0,812.0){\rule[-0.500pt]{1.000pt}{4.818pt}}
\put(676.0,113.0){\rule[-0.500pt]{1.000pt}{4.818pt}}
\put(676,68){\makebox(0,0){3}}
\put(676.0,812.0){\rule[-0.500pt]{1.000pt}{4.818pt}}
\put(828.0,113.0){\rule[-0.500pt]{1.000pt}{4.818pt}}
\put(828,68){\makebox(0,0){4}}
\put(828.0,812.0){\rule[-0.500pt]{1.000pt}{4.818pt}}
\put(980.0,113.0){\rule[-0.500pt]{1.000pt}{4.818pt}}
\put(980,68){\makebox(0,0){5}}
\put(980.0,812.0){\rule[-0.500pt]{1.000pt}{4.818pt}}
\put(1132.0,113.0){\rule[-0.500pt]{1.000pt}{4.818pt}}
\put(1132,68){\makebox(0,0){6}}
\put(1132.0,812.0){\rule[-0.500pt]{1.000pt}{4.818pt}}
\put(1284.0,113.0){\rule[-0.500pt]{1.000pt}{4.818pt}}
\put(1284,68){\makebox(0,0){7}}
\put(1284.0,812.0){\rule[-0.500pt]{1.000pt}{4.818pt}}
\put(1436.0,113.0){\rule[-0.500pt]{1.000pt}{4.818pt}}
\put(1436,68){\makebox(0,0){8}}
\put(1436.0,812.0){\rule[-0.500pt]{1.000pt}{4.818pt}}
\put(220.0,113.0){\rule[-0.500pt]{292.934pt}{1.000pt}}
\put(1436.0,113.0){\rule[-0.500pt]{1.000pt}{173.207pt}}
\put(220.0,832.0){\rule[-0.500pt]{292.934pt}{1.000pt}}
\put(45,472){\makebox(0,0){ $\hbar\omega/\frac{e^2}{\epsilon r_{o}}$}}
\put(828,23){\makebox(0,0){$ql_{o}$}}
\put(1356,742){\makebox(0,0){Fig.2}}
\put(220.0,113.0){\rule[-0.500pt]{1.000pt}{173.207pt}}
\put(235,761){\raisebox{-.8pt}{\makebox(0,0){$\Diamond$}}}
\put(238,698){\raisebox{-.8pt}{\makebox(0,0){$\Diamond$}}}
\put(241,648){\raisebox{-.8pt}{\makebox(0,0){$\Diamond$}}}
\put(248,573){\raisebox{-.8pt}{\makebox(0,0){$\Diamond$}}}
\put(254,518){\raisebox{-.8pt}{\makebox(0,0){$\Diamond$}}}
\put(260,476){\raisebox{-.8pt}{\makebox(0,0){$\Diamond$}}}
\put(273,412){\raisebox{-.8pt}{\makebox(0,0){$\Diamond$}}}
\put(285,365){\raisebox{-.8pt}{\makebox(0,0){$\Diamond$}}}
\put(310,294){\raisebox{-.8pt}{\makebox(0,0){$\Diamond$}}}
\put(335,239){\raisebox{-.8pt}{\makebox(0,0){$\Diamond$}}}
\put(360,191){\raisebox{-.8pt}{\makebox(0,0){$\Diamond$}}}
\put(367,179){\raisebox{-.8pt}{\makebox(0,0){$\Diamond$}}}
\put(373,166){\raisebox{-.8pt}{\makebox(0,0){$\Diamond$}}}
\put(376,159){\raisebox{-.8pt}{\makebox(0,0){$\Diamond$}}}
\put(379,151){\raisebox{-.8pt}{\makebox(0,0){$\Diamond$}}}
\put(382,142){\raisebox{-.8pt}{\makebox(0,0){$\Diamond$}}}
\put(384,137){\raisebox{-.8pt}{\makebox(0,0){$\Diamond$}}}
\put(385,130){\raisebox{-.8pt}{\makebox(0,0){$\Diamond$}}}
\put(387,116){\raisebox{-.8pt}{\makebox(0,0){$\Diamond$}}}
\put(584,117){\raisebox{-.8pt}{\makebox(0,0){$\Diamond$}}}
\put(585,123){\raisebox{-.8pt}{\makebox(0,0){$\Diamond$}}}
\put(589,129){\raisebox{-.8pt}{\makebox(0,0){$\Diamond$}}}
\put(592,133){\raisebox{-.8pt}{\makebox(0,0){$\Diamond$}}}
\put(598,140){\raisebox{-.8pt}{\makebox(0,0){$\Diamond$}}}
\put(610,150){\raisebox{-.8pt}{\makebox(0,0){$\Diamond$}}}
\put(623,157){\raisebox{-.8pt}{\makebox(0,0){$\Diamond$}}}
\put(635,164){\raisebox{-.8pt}{\makebox(0,0){$\Diamond$}}}
\put(660,174){\raisebox{-.8pt}{\makebox(0,0){$\Diamond$}}}
\put(686,181){\raisebox{-.8pt}{\makebox(0,0){$\Diamond$}}}
\put(736,193){\raisebox{-.8pt}{\makebox(0,0){$\Diamond$}}}
\put(786,201){\raisebox{-.8pt}{\makebox(0,0){$\Diamond$}}}
\put(836,207){\raisebox{-.8pt}{\makebox(0,0){$\Diamond$}}}
\put(886,212){\raisebox{-.8pt}{\makebox(0,0){$\Diamond$}}}
\put(936,216){\raisebox{-.8pt}{\makebox(0,0){$\Diamond$}}}
\put(986,220){\raisebox{-.8pt}{\makebox(0,0){$\Diamond$}}}
\put(1036,222){\raisebox{-.8pt}{\makebox(0,0){$\Diamond$}}}
\put(1086,225){\raisebox{-.8pt}{\makebox(0,0){$\Diamond$}}}
\put(1136,227){\raisebox{-.8pt}{\makebox(0,0){$\Diamond$}}}
\put(1186,229){\raisebox{-.8pt}{\makebox(0,0){$\Diamond$}}}
\put(1236,230){\raisebox{-.8pt}{\makebox(0,0){$\Diamond$}}}
\put(1286,232){\raisebox{-.8pt}{\makebox(0,0){$\Diamond$}}}
\put(1336,233){\raisebox{-.8pt}{\makebox(0,0){$\Diamond$}}}
\put(1386,234){\raisebox{-.8pt}{\makebox(0,0){$\Diamond$}}}
\put(1436,235){\raisebox{-.8pt}{\makebox(0,0){$\Diamond$}}}
\end{picture}

\newpage

\setlength{\unitlength}{0.240900pt}
\ifx\plotpoint\undefined\newsavebox{\plotpoint}\fi
\sbox{\plotpoint}{\rule[-0.500pt]{1.000pt}{1.000pt}}%
\begin{picture}(1500,900)(0,0)
\font\gnuplot=cmr10 at 10pt
\gnuplot
\sbox{\plotpoint}{\rule[-0.500pt]{1.000pt}{1.000pt}}%
\put(220.0,113.0){\rule[-0.500pt]{292.934pt}{1.000pt}}
\put(220.0,113.0){\rule[-0.500pt]{1.000pt}{173.207pt}}
\put(220.0,113.0){\rule[-0.500pt]{4.818pt}{1.000pt}}
\put(198,113){\makebox(0,0)[r]{0}}
\put(1416.0,113.0){\rule[-0.500pt]{4.818pt}{1.000pt}}
\put(220.0,233.0){\rule[-0.500pt]{4.818pt}{1.000pt}}
\put(198,233){\makebox(0,0)[r]{1}}
\put(1416.0,233.0){\rule[-0.500pt]{4.818pt}{1.000pt}}
\put(220.0,353.0){\rule[-0.500pt]{4.818pt}{1.000pt}}
\put(198,353){\makebox(0,0)[r]{2}}
\put(1416.0,353.0){\rule[-0.500pt]{4.818pt}{1.000pt}}
\put(220.0,472.0){\rule[-0.500pt]{4.818pt}{1.000pt}}
\put(198,472){\makebox(0,0)[r]{3}}
\put(1416.0,472.0){\rule[-0.500pt]{4.818pt}{1.000pt}}
\put(220.0,592.0){\rule[-0.500pt]{4.818pt}{1.000pt}}
\put(198,592){\makebox(0,0)[r]{4}}
\put(1416.0,592.0){\rule[-0.500pt]{4.818pt}{1.000pt}}
\put(220.0,712.0){\rule[-0.500pt]{4.818pt}{1.000pt}}
\put(198,712){\makebox(0,0)[r]{5}}
\put(1416.0,712.0){\rule[-0.500pt]{4.818pt}{1.000pt}}
\put(220.0,832.0){\rule[-0.500pt]{4.818pt}{1.000pt}}
\put(198,832){\makebox(0,0)[r]{6}}
\put(1416.0,832.0){\rule[-0.500pt]{4.818pt}{1.000pt}}
\put(220.0,113.0){\rule[-0.500pt]{1.000pt}{4.818pt}}
\put(220,68){\makebox(0,0){0}}
\put(220.0,812.0){\rule[-0.500pt]{1.000pt}{4.818pt}}
\put(342.0,113.0){\rule[-0.500pt]{1.000pt}{4.818pt}}
\put(342,68){\makebox(0,0){0.01}}
\put(342.0,812.0){\rule[-0.500pt]{1.000pt}{4.818pt}}
\put(463.0,113.0){\rule[-0.500pt]{1.000pt}{4.818pt}}
\put(463,68){\makebox(0,0){0.02}}
\put(463.0,812.0){\rule[-0.500pt]{1.000pt}{4.818pt}}
\put(585.0,113.0){\rule[-0.500pt]{1.000pt}{4.818pt}}
\put(585,68){\makebox(0,0){0.03}}
\put(585.0,812.0){\rule[-0.500pt]{1.000pt}{4.818pt}}
\put(706.0,113.0){\rule[-0.500pt]{1.000pt}{4.818pt}}
\put(706,68){\makebox(0,0){0.04}}
\put(706.0,812.0){\rule[-0.500pt]{1.000pt}{4.818pt}}
\put(828.0,113.0){\rule[-0.500pt]{1.000pt}{4.818pt}}
\put(828,68){\makebox(0,0){0.05}}
\put(828.0,812.0){\rule[-0.500pt]{1.000pt}{4.818pt}}
\put(950.0,113.0){\rule[-0.500pt]{1.000pt}{4.818pt}}
\put(950,68){\makebox(0,0){0.06}}
\put(950.0,812.0){\rule[-0.500pt]{1.000pt}{4.818pt}}
\put(1071.0,113.0){\rule[-0.500pt]{1.000pt}{4.818pt}}
\put(1071,68){\makebox(0,0){0.07}}
\put(1071.0,812.0){\rule[-0.500pt]{1.000pt}{4.818pt}}
\put(1193.0,113.0){\rule[-0.500pt]{1.000pt}{4.818pt}}
\put(1193,68){\makebox(0,0){0.08}}
\put(1193.0,812.0){\rule[-0.500pt]{1.000pt}{4.818pt}}
\put(1314.0,113.0){\rule[-0.500pt]{1.000pt}{4.818pt}}
\put(1314,68){\makebox(0,0){0.09}}
\put(1314.0,812.0){\rule[-0.500pt]{1.000pt}{4.818pt}}
\put(1436.0,113.0){\rule[-0.500pt]{1.000pt}{4.818pt}}
\put(1436,68){\makebox(0,0){0.1}}
\put(1436.0,812.0){\rule[-0.500pt]{1.000pt}{4.818pt}}
\put(220.0,113.0){\rule[-0.500pt]{292.934pt}{1.000pt}}
\put(1436.0,113.0){\rule[-0.500pt]{1.000pt}{173.207pt}}
\put(220.0,832.0){\rule[-0.500pt]{292.934pt}{1.000pt}}
\put(45,472){\makebox(0,0){$\omega/\omega_{c}$}}
\put(828,23){\makebox(0,0){$ql_{o}$}}
\put(1356,742){\makebox(0,0){Fig.3}}
\put(220.0,113.0){\rule[-0.500pt]{1.000pt}{173.207pt}}
\put(342,233){\raisebox{-.8pt}{\makebox(0,0){$\Diamond$}}}
\put(232,233){\raisebox{-.8pt}{\makebox(0,0){$\Diamond$}}}
\put(244,233){\raisebox{-.8pt}{\makebox(0,0){$\Diamond$}}}
\put(256,233){\raisebox{-.8pt}{\makebox(0,0){$\Diamond$}}}
\put(269,233){\raisebox{-.8pt}{\makebox(0,0){$\Diamond$}}}
\put(281,233){\raisebox{-.8pt}{\makebox(0,0){$\Diamond$}}}
\put(293,233){\raisebox{-.8pt}{\makebox(0,0){$\Diamond$}}}
\put(305,233){\raisebox{-.8pt}{\makebox(0,0){$\Diamond$}}}
\put(317,233){\raisebox{-.8pt}{\makebox(0,0){$\Diamond$}}}
\put(329,233){\raisebox{-.8pt}{\makebox(0,0){$\Diamond$}}}
\put(463,232){\raisebox{-.8pt}{\makebox(0,0){$\Diamond$}}}
\put(706,217){\raisebox{-.8pt}{\makebox(0,0){$\Diamond$}}}
\put(828,206){\raisebox{-.8pt}{\makebox(0,0){$\Diamond$}}}
\put(950,198){\raisebox{-.8pt}{\makebox(0,0){$\Diamond$}}}
\put(1071,191){\raisebox{-.8pt}{\makebox(0,0){$\Diamond$}}}
\put(1193,186){\raisebox{-.8pt}{\makebox(0,0){$\Diamond$}}}
\put(1314,181){\raisebox{-.8pt}{\makebox(0,0){$\Diamond$}}}
\put(1436,177){\raisebox{-.8pt}{\makebox(0,0){$\Diamond$}}}
\sbox{\plotpoint}{\rule[-0.175pt]{0.350pt}{0.350pt}}%
\put(342,334){\makebox(0,0){$+$}}
\put(232,813){\makebox(0,0){$+$}}
\put(244,608){\makebox(0,0){$+$}}
\put(256,517){\makebox(0,0){$+$}}
\put(269,463){\makebox(0,0){$+$}}
\put(281,426){\makebox(0,0){$+$}}
\put(293,398){\makebox(0,0){$+$}}
\put(305,377){\makebox(0,0){$+$}}
\put(317,360){\makebox(0,0){$+$}}
\put(329,346){\makebox(0,0){$+$}}
\put(463,270){\makebox(0,0){$+$}}
\put(706,239){\makebox(0,0){$+$}}
\put(828,237){\makebox(0,0){$+$}}
\put(950,237){\makebox(0,0){$+$}}
\put(1071,237){\makebox(0,0){$+$}}
\put(1193,237){\makebox(0,0){$+$}}
\put(1314,238){\makebox(0,0){$+$}}
\put(1436,238){\makebox(0,0){$+$}}
\end{picture}

\vspace{2cm}
\setlength{\unitlength}{0.240900pt}
\ifx\plotpoint\undefined\newsavebox{\plotpoint}\fi
\sbox{\plotpoint}{\rule[-0.500pt]{1.000pt}{1.000pt}}%
\begin{picture}(1500,900)(0,0)
\font\gnuplot=cmr10 at 10pt
\gnuplot
\sbox{\plotpoint}{\rule[-0.500pt]{1.000pt}{1.000pt}}%
\put(220.0,113.0){\rule[-0.500pt]{292.934pt}{1.000pt}}
\put(220.0,113.0){\rule[-0.500pt]{1.000pt}{173.207pt}}
\put(220.0,113.0){\rule[-0.500pt]{4.818pt}{1.000pt}}
\put(198,113){\makebox(0,0)[r]{0}}
\put(1416.0,113.0){\rule[-0.500pt]{4.818pt}{1.000pt}}
\put(220.0,216.0){\rule[-0.500pt]{4.818pt}{1.000pt}}
\put(198,216){\makebox(0,0)[r]{0.2}}
\put(1416.0,216.0){\rule[-0.500pt]{4.818pt}{1.000pt}}
\put(220.0,318.0){\rule[-0.500pt]{4.818pt}{1.000pt}}
\put(198,318){\makebox(0,0)[r]{0.4}}
\put(1416.0,318.0){\rule[-0.500pt]{4.818pt}{1.000pt}}
\put(220.0,421.0){\rule[-0.500pt]{4.818pt}{1.000pt}}
\put(198,421){\makebox(0,0)[r]{0.6}}
\put(1416.0,421.0){\rule[-0.500pt]{4.818pt}{1.000pt}}
\put(220.0,524.0){\rule[-0.500pt]{4.818pt}{1.000pt}}
\put(198,524){\makebox(0,0)[r]{0.8}}
\put(1416.0,524.0){\rule[-0.500pt]{4.818pt}{1.000pt}}
\put(220.0,627.0){\rule[-0.500pt]{4.818pt}{1.000pt}}
\put(198,627){\makebox(0,0)[r]{1}}
\put(1416.0,627.0){\rule[-0.500pt]{4.818pt}{1.000pt}}
\put(220.0,729.0){\rule[-0.500pt]{4.818pt}{1.000pt}}
\put(198,729){\makebox(0,0)[r]{1.2}}
\put(1416.0,729.0){\rule[-0.500pt]{4.818pt}{1.000pt}}
\put(220.0,832.0){\rule[-0.500pt]{4.818pt}{1.000pt}}
\put(198,832){\makebox(0,0)[r]{1.4}}
\put(1416.0,832.0){\rule[-0.500pt]{4.818pt}{1.000pt}}
\put(220.0,113.0){\rule[-0.500pt]{1.000pt}{4.818pt}}
\put(220,68){\makebox(0,0){0}}
\put(220.0,812.0){\rule[-0.500pt]{1.000pt}{4.818pt}}
\put(423.0,113.0){\rule[-0.500pt]{1.000pt}{4.818pt}}
\put(423,68){\makebox(0,0){2}}
\put(423.0,812.0){\rule[-0.500pt]{1.000pt}{4.818pt}}
\put(625.0,113.0){\rule[-0.500pt]{1.000pt}{4.818pt}}
\put(625,68){\makebox(0,0){4}}
\put(625.0,812.0){\rule[-0.500pt]{1.000pt}{4.818pt}}
\put(828.0,113.0){\rule[-0.500pt]{1.000pt}{4.818pt}}
\put(828,68){\makebox(0,0){6}}
\put(828.0,812.0){\rule[-0.500pt]{1.000pt}{4.818pt}}
\put(1031.0,113.0){\rule[-0.500pt]{1.000pt}{4.818pt}}
\put(1031,68){\makebox(0,0){8}}
\put(1031.0,812.0){\rule[-0.500pt]{1.000pt}{4.818pt}}
\put(1233.0,113.0){\rule[-0.500pt]{1.000pt}{4.818pt}}
\put(1233,68){\makebox(0,0){10}}
\put(1233.0,812.0){\rule[-0.500pt]{1.000pt}{4.818pt}}
\put(1436.0,113.0){\rule[-0.500pt]{1.000pt}{4.818pt}}
\put(1436,68){\makebox(0,0){12}}
\put(1436.0,812.0){\rule[-0.500pt]{1.000pt}{4.818pt}}
\put(220.0,113.0){\rule[-0.500pt]{292.934pt}{1.000pt}}
\put(1436.0,113.0){\rule[-0.500pt]{1.000pt}{173.207pt}}
\put(220.0,832.0){\rule[-0.500pt]{292.934pt}{1.000pt}}
\put(45,472){\makebox(0,0){$\omega/\omega_{c}$}}
\put(828,23){\makebox(0,0){$ql_{o}$}}
\put(1356,742){\makebox(0,0){Fig.4}}
\put(220.0,113.0){\rule[-0.500pt]{1.000pt}{173.207pt}}
\put(230,648){\raisebox{-.8pt}{\makebox(0,0){$\Diamond$}}}
\put(271,685){\raisebox{-.8pt}{\makebox(0,0){$\Diamond$}}}
\put(321,694){\raisebox{-.8pt}{\makebox(0,0){$\Diamond$}}}
\put(326,693){\raisebox{-.8pt}{\makebox(0,0){$\Diamond$}}}
\put(330,692){\raisebox{-.8pt}{\makebox(0,0){$\Diamond$}}}
\put(473,685){\raisebox{-.8pt}{\makebox(0,0){$\Diamond$}}}
\put(524,705){\raisebox{-.8pt}{\makebox(0,0){$\Diamond$}}}
\put(625,740){\raisebox{-.8pt}{\makebox(0,0){$\Diamond$}}}
\put(727,759){\raisebox{-.8pt}{\makebox(0,0){$\Diamond$}}}
\put(828,771){\raisebox{-.8pt}{\makebox(0,0){$\Diamond$}}}
\put(929,780){\raisebox{-.8pt}{\makebox(0,0){$\Diamond$}}}
\put(1031,786){\raisebox{-.8pt}{\makebox(0,0){$\Diamond$}}}
\put(1132,790){\raisebox{-.8pt}{\makebox(0,0){$\Diamond$}}}
\put(1233,794){\raisebox{-.8pt}{\makebox(0,0){$\Diamond$}}}
\put(1335,797){\raisebox{-.8pt}{\makebox(0,0){$\Diamond$}}}
\put(1436,799){\raisebox{-.8pt}{\makebox(0,0){$\Diamond$}}}
\sbox{\plotpoint}{\rule[-0.175pt]{0.350pt}{0.350pt}}%
\put(230,389){\makebox(0,0){$+$}}
\put(271,201){\makebox(0,0){$+$}}
\put(321,134){\makebox(0,0){$+$}}
\put(326,126){\makebox(0,0){$+$}}
\put(329,119){\makebox(0,0){$+$}}
\put(329,120){\makebox(0,0){$+$}}
\put(330,116){\makebox(0,0){$+$}}
\put(473,115){\makebox(0,0){$+$}}
\put(524,140){\makebox(0,0){$+$}}
\put(625,154){\makebox(0,0){$+$}}
\put(727,160){\makebox(0,0){$+$}}
\put(828,163){\makebox(0,0){$+$}}
\put(828,165){\makebox(0,0){$+$}}
\put(929,167){\makebox(0,0){$+$}}
\put(1031,167){\makebox(0,0){$+$}}
\put(1132,168){\makebox(0,0){$+$}}
\put(1233,169){\makebox(0,0){$+$}}
\put(1335,170){\makebox(0,0){$+$}}
\put(1436,170){\makebox(0,0){$+$}}
\end{picture}

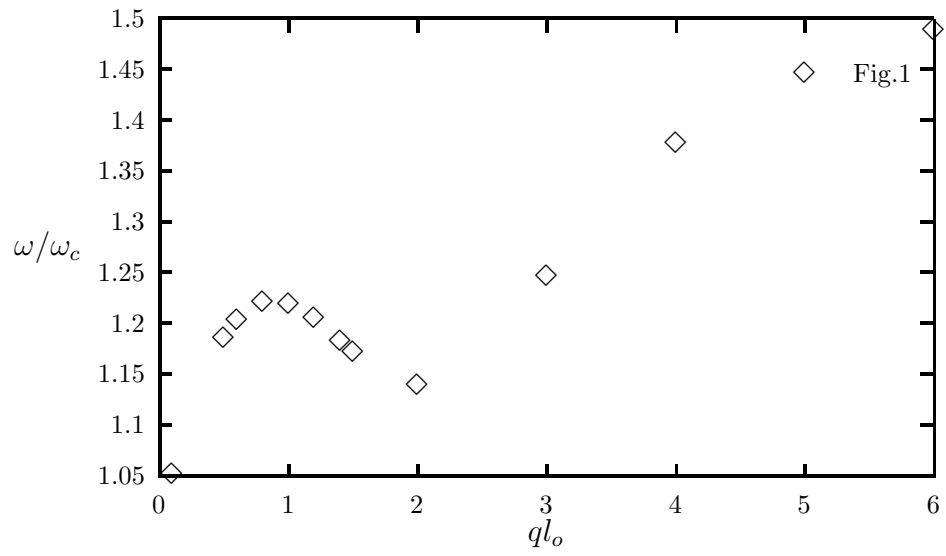
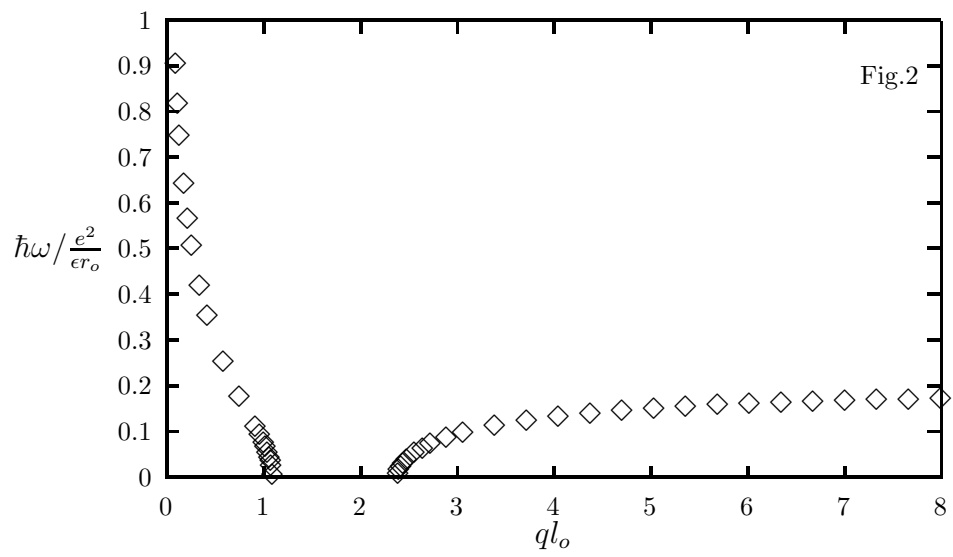
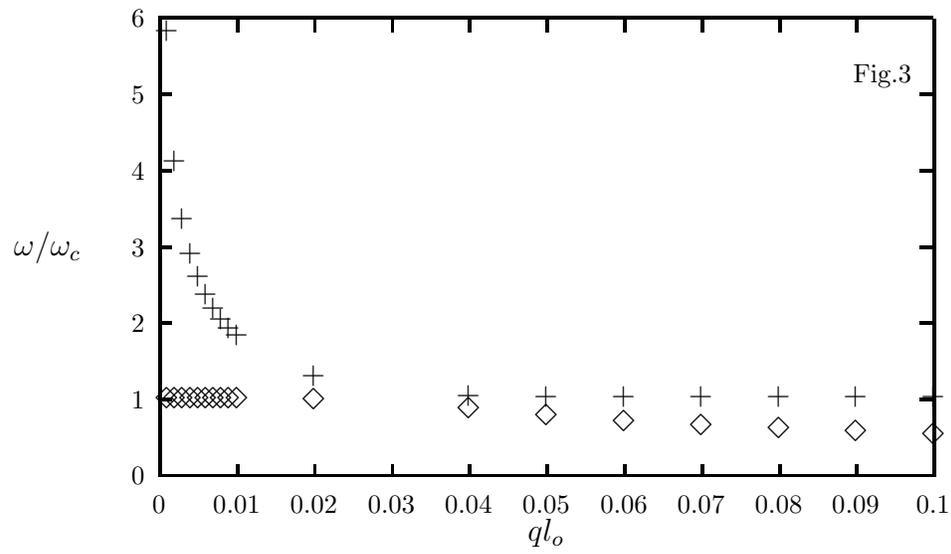
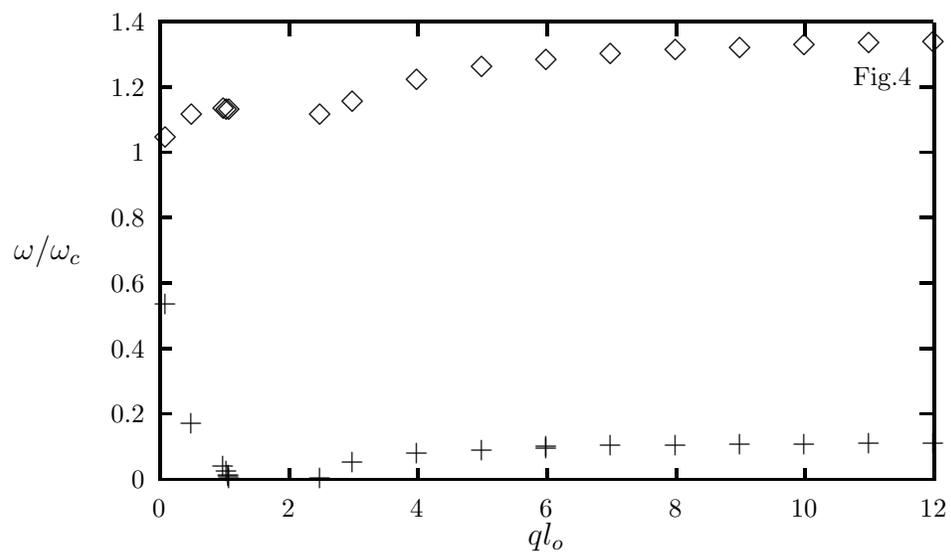
\end{document}